\newcommand{\CLASSINPUTtoptextmargin}{0.75in}
\newcommand{\CLASSINPUTbottomtextmargin}{1.2in}
\documentclass[conference]{IEEEtran}

\IEEEoverridecommandlockouts
\usepackage{cite}
\usepackage{amsmath,amssymb,amsfonts}
\usepackage{nomencl}
\makenomenclature
\usepackage{algorithmic}
\usepackage{graphicx}
\usepackage{textcomp}
\usepackage{xcolor}
\usepackage{subfigure}
\def\BibTeX{{\rm B\kern-.05em{\sc i\kern-.025em b}\kern-.08em
    T\kern-.1667em\lower.7ex\hbox{E}\kern-.125emX}}
\usepackage[utf8]{inputenc}
\begin{document}

\title{Toward a Mathematical Vulnerability Propagation and Defense Model in Smart Grid Networks\\
}

\author{Abhijeet Sahu,~\IEEEmembership{Student Member,~IEEE}, Bin Mai, 
Katherine Davis,~\IEEEmembership{Senior Member,~IEEE}, Ana Goulart

}


\maketitle

\begin{abstract}
For reducing threat propagation within an interconnected network, it is essential to distribute the defense investment optimally. Most 
electric power utilities are resource constrained, yet how to account for costs while designing threat reduction techniques is not well understood.
Hence, in this work, a vulnerability propagation and a defense  model is proposed based on an epidemic model. 
The new defense mechanism is then validated through sensitivity of the propagation parameters on the optimal investment with two-node and N-node cases. Further, the model efficacy is evaluated  with implementation in one of the communication networks of a cyber-physical power system. Topological impact on the optimal nodal investment is also emphasized. 
Optimal investment of the neighbors with less degree were 
found to be 
highly
sensitive to fluctuation in vulnerability exploitability probability.
\end{abstract}



\nomenclature{$v_i$}{Probability of vulnerability exploitation at $i$}%
\nomenclature{$\alpha_{ij}$}{Vulnerability propagation factor from $i$ to $j$}
\nomenclature{$v_i^{'}$}{$v_i$ update after single step propagation.}
\nomenclature{$\gamma$}{Multiplicative factor}
\nomenclature{$\theta$}{Inverse power factor}
\nomenclature{$z_i$}{Investment at node $i$}
\nomenclature{$W$}{Overall budget of the firm}
\nomenclature{$\hat{v_i}$}{$v_i$ update after first defense action.}
\nomenclature{$\bar{v_i}$}{$v_i$ update after second step propagation after defense.}
\printnomenclature

\section{Introduction}

In a communication network consisting of multiple inter-connected nodes, one significant element of the security of the network is that one node's security and vulnerability level has impacts on the security of other logically adjacent nodes.
Cyber attacks such as ransomware or botnets impact the communication devices where these malicious tools are installed, then can expand to the devices and communication nodes that are in the same network.  The recent distributed Ransomware-As-A-Service attack on the Colonial Pipeline in May 2021 was multi-stage, accomplished by a set of affiliates working together to exploit vulnerabilities at different parts of the
network. For instance, one affiliate UNC2659 exploited vulnerability 
CVE-2021-20016
in the SonicWall firewall and installed the TeamViewer application on the protected network to get remote access, followed by utilization of the Beacon botnet to run the Mimikatz credential theft tool by another affiliate UNC2628, to finally install the SMOKEDHAM backdoor~\cite{darkside}. Developing a mathematical model of \textit{vulnerability propagation (VP)} in cyber-physical control networks will assist in prioritizing defense investment to prevent such multi-stage attacks.  

There are numerous VP models proposed in literature. 
For example, the VP model in \cite{vp_1} for distributed virtualized systems is based on inter-clusters, with the same vulnerability in a cluster, utilizing ordinary differential equations to discover the law of VP, while~\cite{vp_2} provides an adaptive ecosystem-inspired model of vulnerability exploitation and risk flow in a heterogeneous network using a multi-agent epidemic model. 
Though these propagation models incorporate multi-stage threats and their integration in a distributed virtualized environment, they have not studied the impact of the model parameters to the defense methodologies through rigorous sensitivity analysis.
Moreover, the existing models also do not address how to determine optimal cyber investment to address the impact of such threats. Few works, such as~\cite{optimal_investment} and~\cite{optimal_investment2}, develop  economic models for an optimal level of information security investment. This work takes the benefit of both the existing epidemic propagation models and the optimized defense investment models, and proposes a novel VP model that is integrated with 
the optimal investment problem, and it validates the solution through sensitivity analysis.





To defend the network against these attacks, an information technology (IT) department has to 
invest on defense mechanism such as firewalls, Security Information and Event Management (SIEMS), Anti-Virus servers, vulnerability and software patches, etc.
This is an organization's
\emph{security investment}.
The investment is 
based on 
security budget $W$ which is limited based on the firm's policy. The goal of this paper is to show the following:

\begin{itemize}
    \item How to minimize the vulnerability of the network nodes, i.e., to increase the resilience of the network, by formulating an optimization problem with
    constraint 
    that cost to mitigate the security vulnerability is within
    budget $W$.
    \item Evaluate the sensitivity of the optimal investment by varying VP parameters.
    \item Integrate the model for varying network topologies of synthetic electric power utilities, with extension to different Industrial Control System (ICS) networks.
\end{itemize}

In summary, the research question is to 
develop and validate a model of VP in smart grid networks
that 
also 
guides
optimal cybersecurity investment 
in utilities 
toward optimal vulnerability mitigation, 
while satisfying an organization's
real-world policies and budget.
Then, this model is incorporated for a cyber-physical power system case, adopted from the synthetic Texas CP-2000 model~\cite{Wlazlocp2000}. 


The paper proceeds as follows. Section~\ref{proposed_model} presents the proposed model for optimal investment under VP and reviews prior work.
Section~\ref{model_definition} defines the steps of the model. Section~\ref{model_analysis} presents the analysis of the model with a two node case, sensitivity analysis, and generalizes the solution for $N$ node scenario. Finally, Section~\ref{results} validates the 
efficacy for two nodes, $N$ nodes, and two synthetic
power system networks,
and Section~\ref{conclusion} concludes the paper. 

\section{Proposed Model}\label{proposed_model}

In the proposed model, a communication network is considered of $n$ nodes, each denoted by
$a_i$ for $(i = 1, ...n)$. 
A \textit{default vulnerability}, $v_i~\epsilon~[0,1]$ is defined for each node $a_i$, 
as the probability that any one of the vulnerability on a node $a_i$, is successfully exploited by the intruder.
In this case, \emph{default} means the value of $v_i$, when the node 
is isolated i.e. it does not propagate the vulnerability further. 
The reason for defining a default vulnerability is that to model VP, one need an initial vulnerability state, and these values help give us an estimate of initial `state'.
In this work, it is assumed
that this value can be computed, i.e., based on the average of the exploitability subscore obtained from the Common Vulnerability Scoring System (CVSS) values in the National Vulnerability Database (NVD). Dynamic update of $v_i$ is made based on the vulnerabilities found in the node, as well as any updates based on the defense investment. 

When multiple nodes are connected to form a communication network, then any node's vulnerability will have an impact on the adjacent nodes connected to it. In our model, when a node $a_i$ has been successfully attacked, 
it 
activates what is defined as a \textit{VP factor}, $\alpha_{ij}~\epsilon~[0,1]$, that increases the value of $v_j$, the default vulnerability of node $a_j$ that is connected to $a_i$. A lower value is more severe. As the value of $\alpha_{ij}$ increases, the 
VP between nodes $a_i$ and $a_j$ has less impact. If the value of $\alpha_{ij}$ is exactly equal to 0, then $a_j$ will be completely vulnerable once $a_i$ has been successfully attacked, or in other words, an attack on $a_j$ is guaranteed to succeed. If the value of $\alpha_{ij}$ is exactly equal to 1, then a attack on $a_i$ will not affect the vulnerability of $a_j$. After propagation occurs to node $a_j$, its vulnerability becomes $v_j$ = $v_j^{\alpha_{ij}}$. The rationale behind considering an exponential function is based on a prior epidemic model proposed 
in various worm propagation models such as Uniform Scan Worm Model~\cite{worm_prop1}, where the infected number of hosts is exponential with respect to time, and the Code Red Worm propagation following the classical epidemic Kermack-Mckendrick model~\cite{code_red}. Even the spreading pattern of the mobile phone viruses~\cite{mobile_virus_prop} follows a similar epidemic model. Unlike these models
where the net infected hosts increase exponentially with time, we propose a novel fine-grained
model which updates individual node's vulnerability exponentially depending on the propagation of vulnerability from the neighbor nodes. 
The VP factor between two nodes is asymmetric, i.e. $\alpha_{ij} \neq \alpha_{ji}$. The intuition for the asymmetric nature is due to the fact that different vulnerabilities existent in the two nodes at a given time. 

Most of the prior models on defense that have been proposed follow a system-wide 
time-dependent node recovery model. For instance, authors in~\cite{quarantine} propose a Dynamic Quarantine Defense model which tries to reduce the number of infected nodes system-wide. While this work proposes a Security Investment based budget model specific to a node, where a defender invest in node $a_i$ to lower its $v_i$ value, thus reducing the probability of an attack being successful. After a investment amount of $z_i$, $v_i$ is lowered according to the following equation $\hat{v_i} = \frac{v_i}{(\gamma v_i+1)^\theta}$ where $\gamma$ $>$ 0 and $\theta$ $\geq$ 1. The total security budget is referenced by the variable $W$.

\section{Model Definition}\label{model_definition}
In a real network, it is unrealistic to obtain an equilibrium vulnerability, where the vulnerability probability saturates to a fixed value, since the intrusion and defense actions never stops ideally. In the proposed model it is assumed an equilibrium is obtained in two stages of VP, hence the analysis would involve four steps to find the
the final equilibrium vulnerability
value at each node,
which is denoted as a node's value $\bar{v_i}$ after all defenses, when vulnerabilities are no longer changing.

\begin{enumerate}
\item Obtains the default vulnerability of each individual node: $v_i$ for $(i = 1, ...n)$. It is assumed that the values of $v_i$ are provided exogenously. 

\item Finds the vulnerability value $v_i'$ of each node, after VP within the network, using the following: 
\begin{equation}\label{step1}
    v_i' = v_i^{(v_j'\alpha_{ji}+1-v_j')}
\end{equation}

\item Computes $v_i'$, the individual vulnerability after a security investment $z_i$ has been made, using the following, 
\begin{equation}
    \hat{v_i} = \frac{v_i'}{(\gamma z_i+1)^\theta}
\end{equation}
where $\gamma$ and $\theta$ are 
constants that represent the multiplicative and exponential investment factors, respectively.
Factor $\gamma$ indicates the effectiveness of the investment in reducing the vulnerability, and $\theta$ indicates the defense reinforcement (higher $\theta$ reinforces the defense to reduce vulnerability). 


\item In last step, all the nodes in the network, after security investment, have their vulnerability propagated again, which results in the final equilibrium vulnerability as: 
\begin{equation}
\bar{v_i} = \hat{v_i}^{\prod(\bar{v_j}\alpha_{ji}+1-\bar{v_j})}
\end{equation}
\end{enumerate}

Therefore, the objective function of the model is: $min \sum_{i=1}^{n} \bar{v_i}$. The constraint is $\sum_{i=1}^{n}{z_i} \leq {W}$. It is an optimization problem with a nonlinear relationship between the 
variables, as well as a nonlinear objective. As the number of nodes in the network increases, solving the optimization problem is computationally more expensive, hence we linearize the exponential relations
using the Taylor Series:
\begin{equation}
a^{x}=e^{x \ln a}=1+\frac{x \ln a}{1 !}+\frac{(x \ln a)^{2}}{2 !}+\frac{(x \ln a)^{3}}{3 !}+\cdots
\end{equation}
By considering the first order Taylor series approximation, the second step of the model is linearized as follows:
\begin{equation}
    v_i' = 1 + (v_j'\alpha_{ji} + 1 - v_j')log_{n}v_i
\end{equation}

Among the four steps described in the model, for ease of exposition of the work and to build intuition about how the proposed models work, first a \textbf{simple model} is considered that involves only the first three steps, 
i.e.,
a single step 
in the VP. Further, the model is analyzed, considering the fourth step with a double step in VP, called the \textbf{two-stage model}. For the \textbf{simple model} the objective to minimize is sum of $min \sum_{i=1}^{n} \hat{v_i}$ and for the \textbf{two-stage model} it is $min \sum_{i=1}^{n} \bar{v_i}$. The model and the defense framework are then generalized for $n$-step attacks. 
To validate the efficacy of the model, multiple sensitivity analyses are performed, such as the impact on the optimal $z_i$ with varying  $\alpha$, $v$, and $W$.

\section{Model Analysis}\label{model_analysis}

\subsection{Two-Node Case}
In the two-node case considered, both nodes are able to affect each other when an attack may happen on either node.
The two nodes in this model are $a_1$ and $a_2$. Node $a_1$ has its own default vulnerability, $v_1$, its security investment, $z_1$, and a VP factor, $\alpha_{12}$. Similarly, node $a_2$ has 
$v_2$, $z_2$, and $\alpha_{21}$. In this model, the decision variables are $z_1$ and $z_2$, while all other values are parameters.  The four model steps are applied:

\begin{enumerate}
    \item Apply the parameters previously mentioned relating to $a_1$ and $a_2$. 
\item 
Utilize the equation, $v_i'$ = $v_i^{(v_j'\alpha_{ji}+1-v_j')}$, to find the vulnerability value after propagation for both nodes. 
In this case, the equations for $v_1'$ and $v_2'$ correspond to the following:
\begin{gather*}
a_1,\ v_1' = v_1^{(v_2'\alpha_{21}+1-v_2')} \;\;\;
a_2,\ v_2' = v_2^{(v_1'\alpha_{12}+1-v_1')}
\end{gather*}
Solving these equations would result in 
the symbolical results of $v_1'$ and $v_2'$.

\item Computes the $\hat{v_i}$, the individual vulnerability after a security investment of 
$z_i$, following these equation: $\hat{v_i} = \frac{v_i'}{(\gamma v_i+1)^\theta}$. The corresponding equations for a two-node model are:
\begin{gather*}
a_1,\ \hat{v_1} = \frac{v_1'}{(\gamma v_1+1)^\theta}\;\;\;
a_2,\ \hat{v_2} = \frac{v_2'}{(\gamma v_2+1)^\theta}
\end{gather*}

\item The last step is similar to the second step 
with a few slight differences. The process replaces $v_i$ with $\hat{v_i}$ and $v_i'$ with $\bar{v_i}$. The corresponding equations for this two-node model are as follows:
\begin{gather*}
a_1,\ \bar{v_1} = \hat{v_1}^{(\bar{v_2}\alpha_{21}+1-\bar{v_2})}\;\;\;
a_2,\ \bar{v_2} = \hat{v_2}^{(\bar{v_2}\alpha_{12}+1-\bar{v_2})}
\end{gather*}
This step solves for $\bar{v_1}$ and $\bar{v_2}$ symbolically as expressions of $z_1$ and $z_2$.
\end{enumerate}

The ending objective function of the two-node model is $min\ \bar{v_1} + \bar{v_2}$. The constraint is $z_1$ + $z_2$ $\leq$ $W$. $W$ in this case is a parameter. Incorporating the linearization discussed in the previous section for the \textbf{simple model}, the $\bar{v_1}$ and $\bar{v_2}$ is computed in terms of  $v_1$, $v_2$, $\gamma$, $\theta$, $\alpha_{12}$, and $\alpha_{21}$, resulting in the following
objective function $\bar{v_1} + \bar{v_2}$:
\begin{equation}\label{eqn:v1pv2}
    \frac{2 + b_1(\alpha_{21} + b_2(\alpha_{21} - 1)) + b_2(\alpha_{12} + b_1(\alpha_{12} -1))}{ 1- b_1 b_2(\alpha_{12} - 1)(\alpha_{21} - 1)}
\end{equation}
In Eq.~\ref{eqn:v1pv2}, $b_i$ is given by $\log(v_i) - \theta(\log(\gamma\;z_i) + 1)$.
For the \textbf{two-stage model}, $b_i$ is given by $\log(k_i/k) - \theta(\log(\gamma z_i) + 1)$, where $k_i = 1 + \log{v_i}\alpha_{ji} + (\alpha_{ji} - 1)\log{v_i}\log{v_j}$ and $k = 1 - (\alpha_{ji}- 1)(\alpha_{ij}- 1)\log{v_i}\log{v_j}$

\subsection{Sensitivity Analysis}\label{sensitivity_analysis}
The sensitivity analysis of $v_i$ for node $i$
with respect to the investment $z_i$, is analytically computed for the two-node case. The $\frac{\mathrm{d \bar{v_i}}}{d z_i}$ obtained is proportional to the following,
\begin{equation}\label{sensitivity_1}
    \frac{(\log{v_i} - \theta \log(\gamma z_i + 1)) \theta k \gamma}{(1 + ak')^2 (\gamma z_i + 1)}
\end{equation}
where $k = b(\alpha_{12} - 1)(\alpha_{21} - 1)(b(\alpha_{12}(\alpha_{21} - 1)) - \alpha_{21})$ and $k' = b(\alpha_{12} - 1)(\alpha_{21} - 1)$. The expressions $a$ and $b$ refer to $\log{v_1} - \theta \log(\gamma z_1 + 1)$ and $\log{v_2} - \theta \log(\gamma z_2 + 1)$, respectively. 

Eq.~\ref{sensitivity_1}'s value indicates how much 
a node's vulnerability probability
reduces with rise in investment. A higher $\theta$ and $z_i$ 
reduce Eq.~\ref{sensitivity_1} as well as $v$. 
This is a second-order effect: adding security investment actually reduces the node's sensitivity - it also means that the initial state is not near the optimal point, and that the optimization can work as small changes in investments are capable of driving the sensitivity closer to zero, to obtain an equilibrium in multiple stages of propagation.
Analyzing further, to understand how an optimal $z_i^{*}$ is affected by an initial vulnerability $v_i$, Eq.~\ref{sensitivity_1} is set to 0 and solved to obtain $z_i^{*} = \frac{v_i^{\theta^{-1}} - 1}{\gamma}$. 
This 
indicates that optimal investment required increases with rise in vulnerability.

\subsection{N-Node Generalization}

Building on the basics
of the two-node case, an N-node case is considered next. With the \textbf{simple model}, $\hat{v_i}$ in the objective function is trivial and computed as follows:
\begin{equation}\label{eq_ref1}
    \hat{v_i} = 1 + \log{v_i'}\prod_{j}{}\hat{v_j}(\alpha_{ji} -1) + 1 
\end{equation}

By comparison, obtaining $\bar{v_i}$, in the N-node scenario for the  \textbf{two-stage model} is non-trivial. The values of $\bar{v_i}$ for each node are computed as follows,

\begin{equation}\label{eq_ref1}
    \bar{v_i} = 1 + \log(\hat{v_i})\prod_{j}{}\bar{v_j}(\alpha_{ji} -1) + 1 
\end{equation}
where $\log(\hat{v_i})$ is computed as:
\begin{equation}\label{eq_ref1}
    \log(\hat{v_i}) = \prod_{i}{}\log{v_i'} - \theta \log(\gamma z_i + 1)
\end{equation}

The challenge exists in computing the $v_i'$ analytically for a N-node system. Newton's method is utilized to obtain $v_i'$ through an iterative method which 
converges when $|v_{i, n+1}' - v_{i,n}'| \le T$ with threshold $T$,

\begin{equation}
    v_{i, n+1}' = v_{i,n}' - \frac{f(v_{i,n}')}{f'(v_{i,n}')}
\end{equation}
where $f$ is a function based on Eq.~\ref{step1}, and
$f'$ is the Jacobian of $f$ which is a 2-dimensional matrix, with diagonal elements all 0. The non-diagonal elements are non-zero based on the existence of an edge between node $i$ and $j$, and  $v_{i,n}'$ denotes the $v_i'$ at $n^{th}$ iteration.

The non-zero values of the Jacobian $f'$ are computed as,
\begin{equation}
    f'_{ij} = 1 + \sum_{p=1}^{c(K)} \prod_{l=1}^{c(K_p)} v_{K_{pl}} (\alpha_{K_{pl}i} - 1)
\end{equation}
where, $K$ is the combination of all the pairs of number possible, excluding $i$ and $j$, and nodes excluding neighbors of $a_i$. For instance, for $n=4$ with $i=1$ and $j=2$, the set $K$ will be $\{3,4,34\}$, assuming both node $a_3$ and $a_4$ are connected to node $a_1$ . The term $c(K)$ refers to the cardinality of the set $K$, which is 3 for the set $\{3,4,34\}$. $K_p$ refers to each element in $K$ indexed through $p$, hence $K_p$ for $p=3$ is the third element in $K$, $34$, $c(K_p)$  refers to the cardinality of the number of nodes in $K_p$, hence $c(K_p) = 2$, i.e nodes $a_3$ and $a_4$. $K_{pl}$ refers to the node index, i.e. for $p=3,l=1$, $K_{pl}$ is $3$, while for $p=3,l=2$, $K_{pl}$ is $4$. 
$\alpha_{K_{pl}i}$ is the VP factor between node $K_{pl}$ and $i$.

\section{Results \& Analysis}\label{results}
In this section, first the VP model is evaluated for the two node case with the simple and two-stage scenarios. Further, the N-node scenario is considered, with different network topologies, and the model is implemented
on two cyber-physical power system networks from a 
synthetic communication network for a 2000-bus synthetic electric grid case~\cite{Wlazlocp2000}. For all the experiments, the parameters $\theta$ and $\gamma$ were kept fixed at $2$ and $0.7$, respectively. The source codes of the model and the optimization problem solution are available in Github~\cite{github_vuln_prop}.
\subsection{Two Node Case}


\subsubsection{Simple Model}
\begin{figure}[h]
 \centering
     \includegraphics[width=0.9\columnwidth]{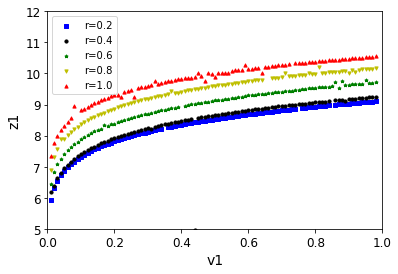}
     \caption{\textbf{Simple Model:} Impact of varying vulnerability probability $v_1$ on the optimal $z_1^{*}$, evaluated with different $r$}
     \label{fig:simple_2n_v_z}
     \vspace{-4mm}
 \end{figure}

\textit{a) Impact of $v$ and $\alpha$:} Fig.~\ref{fig:simple_2n_v_z} shows the effect of altering $v_1$ on the optimal investment $z_1$. The $r$ (asymmetric factor) in the figure refers to the ratio of the VP factor among the two nodes, i.e., $r=\frac{\alpha_{12}}{\alpha_{21}}$. The optimal investment increases with $v_1$ as well as with an increase in $r$, indicating that if the source of intrusion has higher potential for propagation compared to its destination, then a major investment is required at the source. 
The \textit{Two Node} case is considered to show empirically and analytically, that the sensitivity follows an exponential relationship.

\begin{figure}[h]
 \centering
     \includegraphics[width=0.9\columnwidth]{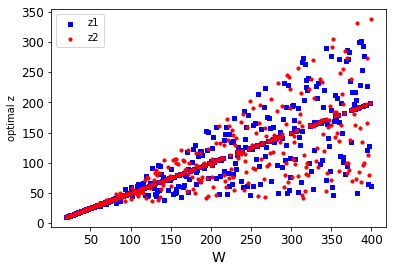}
     \caption{\textbf{Simple Model} Impact of $W$ on the optimal $z$ with fixed $v$'s and $\alpha$'s. 
     }
     \label{fig:impact_w_2n}
     \vspace{-4mm}
 \end{figure}

 \textit{b) Impact of net investment $W$:}
    Fig.~\ref{fig:impact_w_2n} shows the effect of $W$ on the optimal investment $z$, which increases, but there is a lot of variance after a certain budget level is reached. This experiment is carried out while holding values for $v$ and $\alpha$ constant, i.e., 0.5.

\subsubsection{Two-Stage Model}

\begin{figure}[h]
 \centering
     \includegraphics[width=0.9\columnwidth]{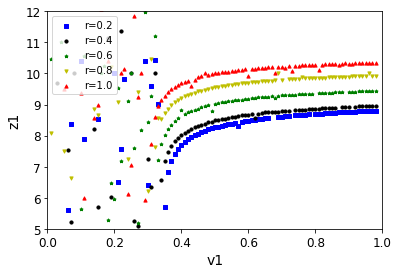}
     \caption{\textbf{Two-Stage Model:} Impact of varying vulnerability probability $v_1$ on the optimal $z_1^{*}$. Evaluated with different $r$ is the ratio $\frac{\alpha_{12}}{\alpha_{21}}$}
     \label{fig:impact_2n_v_z_c}
 \end{figure}

\textit{a) Impact of $v$ and $\alpha$:}
    Fig.~\ref{fig:impact_2n_v_z_c} shows the two-stage model results, where an interesting observation can be made. For $v_1$ less than $0.35$, the optimal $z_1$ is quite random. The probable reason may be that a lower initial $v$, with subsequent attacks, 
    causes much greater deviation in $\bar{v_1}$ which forms the objective function.  Higher $r$ makes the investment higher for node $a_1$, as the attack on node $a_2$ is high due to higher $\alpha_{12}$. 
 
 \textit{b) Impact of net investment $W$:} 
 The two-stage model shows a similar response to the simple model, when analyzing the impact of net investment $W$ on optimal $z$. 

\subsection{N-Node Case}

\subsubsection{Simple Model}
 


\begin{figure*}[!htb]
  \centering
  \subfigure[]{\includegraphics[width=0.32\textwidth]{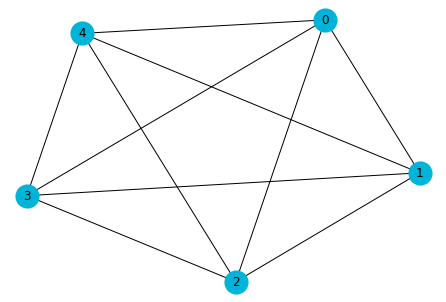}\label{dense_graph}}
  \subfigure[]{\includegraphics[width=0.32\textwidth]{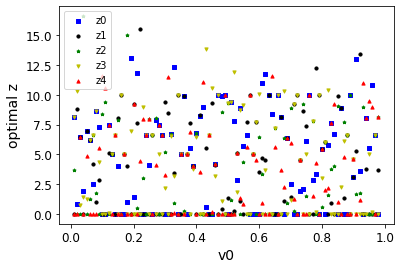}\label{dg_01}}
  \subfigure[]{\includegraphics[width=0.32\textwidth]{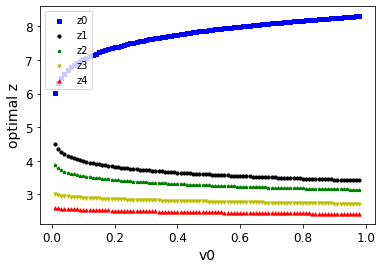}\label{dg_08}}
  \caption{Impact of varying vulnerability probability $v_0$ on the optimal $z^{*}$ with (a) Dense Graph (b) $\alpha$'s = 0.1 (c) $\alpha$'s = 0.8}
\end{figure*}

\begin{figure*}[!htb]
  \centering
  \subfigure[]{\includegraphics[width=0.32\textwidth]{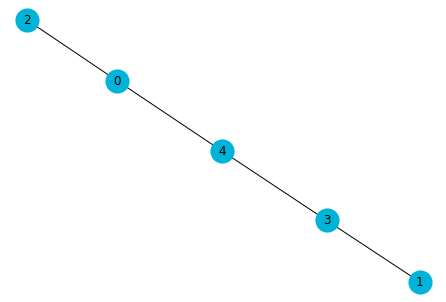}\label{sparse_graph}}
  \subfigure[]{\includegraphics[width=0.32\textwidth]{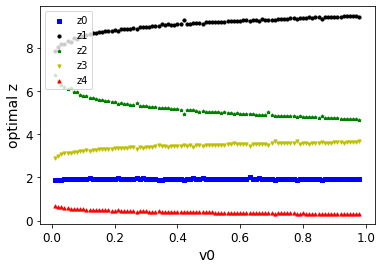}\label{sg_01}}
  \subfigure[]{\includegraphics[width=0.32\textwidth]{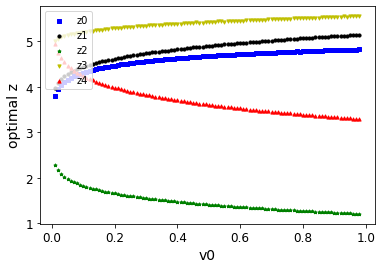}\label{sg_08}}
  \caption{Impact of varying vulnerability probability $v_0$ on the optimal $z^{*}$ with (a) Sparse Graph b) $\alpha$'s = 0.1 (c) $\alpha$'s = 0.8}
\end{figure*}

The effect of network topology is studied for a dense and sparse graph with $N=5$.
    Fig.~\ref{dense_graph} and Fig.~\ref{sparse_graph} are the dense and sparse graphs, respectively.
    Figs.~\ref{dg_01} and~\ref{dg_08} show the impact of varying  vulnerability probability $v_0$ on the optimal $z^{*}$ for $\alpha$'s = 0.1 and $\alpha$'s = 0.8, respectively. It can be observed that for lower $\alpha$, the optimal solutions have higher variance. Figs.~\ref{sg_01} and~\ref{sg_08} show the impact study for the sparse graph.
    In the case of the dense graph, with an increase in $v_0$, most nodes' optimal investment  decreased for other nodes except itself. For the sparse graph, however, immediate neighbors' had their optimal cost reduced, while optimal cost increased for more distant nodes. 
    The probable reason for such observation, is the assumption that all the nodes at the beginning of the propagation hold same $v$ and the net investment is fixed at $W$. In a dense graph, as a higher degree node's $v$ increases, the impact is distributed among all the neighbors. So the decrease of optimal cost in a neighboring node is lesser in comparison to a neighboring node in a sparse graph with lower degree nodes.   

    
\subsubsection{Evaluation in a Cyber-Physical System}

A large-scale
cyber-physical synthetic electric grid model on the footprint of Texas from~\cite{Wlazlocp2000} is considered, where the cyber components of this power system case have been created following the communication model adopted by a regional reliability coordinator 
in interacting with the electric utilities or the market participants (i.e., a scheduling entity (QSE), load-serving entity (LSE), resource entity (RE), or transmission/distribution service provider (TDSP)). The communication model is hierarchical based on three primary levels: Level 1: Balancing Authorities,  Level 2: Utility Control Center (UCC), and  Level 3: Substations. 
In this work, the VP of one utility, $Utility\_0$ and a substation $Odessa2$ have been analyzed, from the synthetic model~\cite{Wlazlocp2000}.

\begin{figure}[h]
 \centering
     \includegraphics[width=0.9\columnwidth]{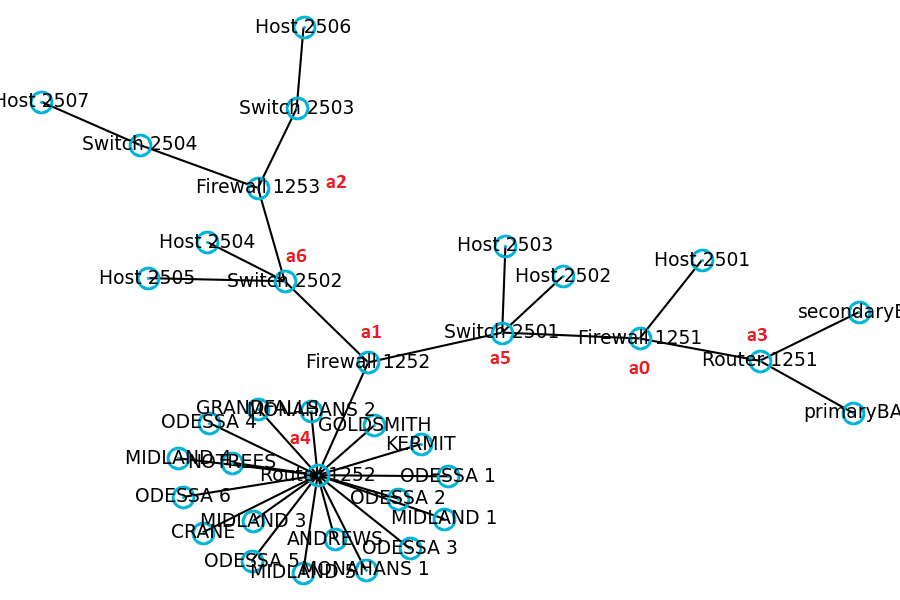}
     \caption{$Utility\_0$ connected to their substations alongwith its internal components connected to the primary and secondary Balancing Authority (BA) in the right.}
     \label{fig:util0}
 \end{figure}

\begin{figure*}[!htb]
  \centering
  \subfigure[]{\includegraphics[width=0.32\textwidth]{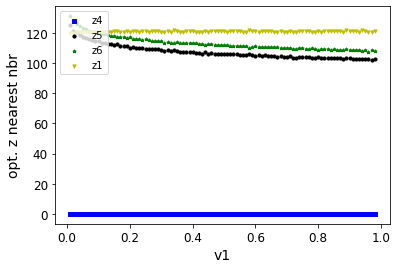}\label{a1}}
  \subfigure[]{\includegraphics[width=0.32\textwidth]{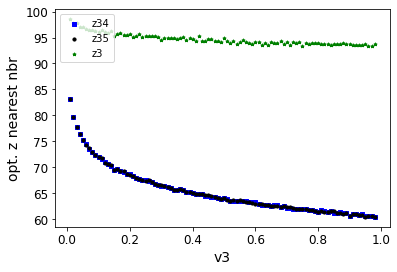}\label{a3}}
  \subfigure[]{\includegraphics[width=0.32\textwidth]{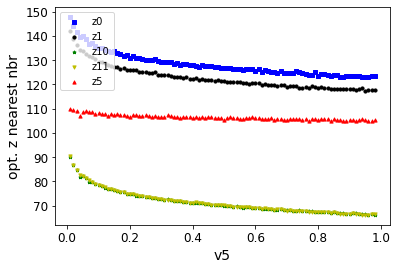}\label{a5}}
  \caption{Impact of varying vulnerability probability $v$'s on the optimal $z^{*}$ on its neighboring node with $v$ modified at the nodes a) $a_1$, a Firewall, b) $a_3$, a Router, and c) $a_5$, a Network Switch}
\end{figure*}

\textbf{Utility Control Center}:
    A UCC hosts multiple servers for dedicated purposes such as its Energy Management System (EMS) and applications
    such as state estimation, transient stability, small signal stability, voltage stability, and post-fault analysis.
   Fig.~\ref{fig:util0} shows the graph for a UCC, $Utility\_0$, with 19 connected substations. 
   From the concept of cyber-physical betweenness centrality theory of an attack graph in an ICS network~\cite{cpbc}, protecting a higher degree node reduces intruders attack paths.
   Hence, the VP model is analyzed by picking nodes with at least 3-degree.
   Figs.~\ref{a1}-\ref{a5} indicate the effect of vulnerability probability of a few selected nodes based on their degree and type, on the neighboring node's optimal cost of investment. Fig.~\ref{a1} shows that varying $v_1$ effects the $z^{*}$ more for the neighbors with less degree, i.e., $a_6$ and $a_5$.

\begin{figure*}[!htb]
  \centering
  \subfigure[]{\includegraphics[width=0.32\textwidth]{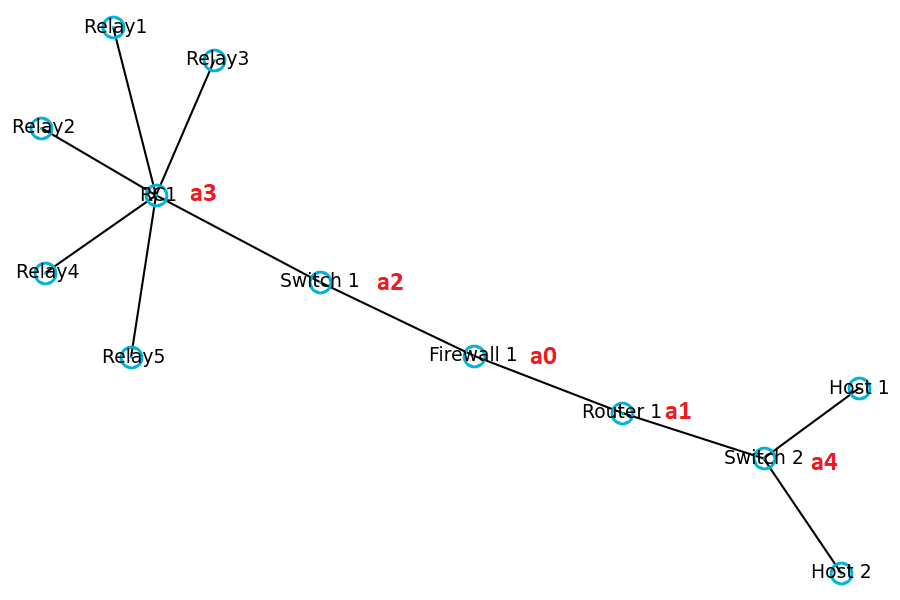}\label{s0}}
  \subfigure[]{\includegraphics[width=0.32\textwidth]{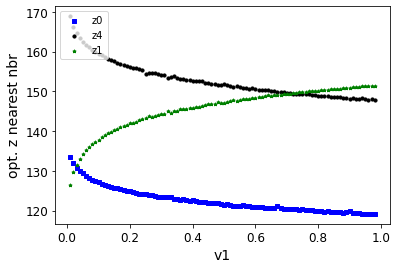}\label{s2}}
  \subfigure[]{\includegraphics[width=0.32\textwidth]{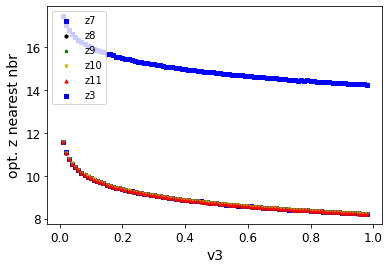}\label{s4}}
  \caption{a) $Odessa\_2$ substation network connected to $Utility\_0$. b) Impact of varying vulnerability probability $v$'s on the optimal $z^{*}$ on its neighboring node with $v$ modified at the nodes $a_1$ and c) $a_3$}
\end{figure*}

\textbf{Substation}: A substation consists of a local control center with devices stationed in three different levels. At the station level, the operator workstation, controllers, DNP3 outstations, local DB or web server are deployed. At the bay level, 
remote terminal units (RTUs), bay controllers, and relays 
are deployed. At the process level, current transformers (CTs), potential transformers (PTs), circuit breakers, etc., are directly connected to feeders or transformers. Fig.~\ref{s0} shows the graph for a substation, $Odessa\_2$. 
The VP model is analyzed, picking the nodes with at least 2-degree, represented through nodes $a_0$ through $a_5$ in the Fig.~\ref{s0}. Figs.~\ref{s2}-\ref{s4} indicate the effect of altering $v$ of a relay controller ($a_3$) and a router ($a_1$), on their neighboring node's $z^{*}$. A relay controller with higher node degree have same effects on its neighbor (Fig.~\ref{s4}), while the router with lower degree have non-equal effect on its neighbor (Fig.\ref{s2}), primarily depending on the neighbor node's degree. 
    
\subsubsection{Two-Stage Model}

\textit{Impact of Network Topology:}

\begin{figure*}[!htb]
  \centering
  \subfigure[]{\includegraphics[width=0.32\textwidth]{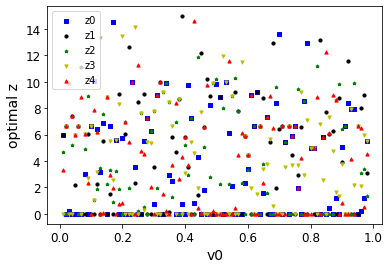}\label{dg_01_c}}
  \subfigure[]{\includegraphics[width=0.32\textwidth]{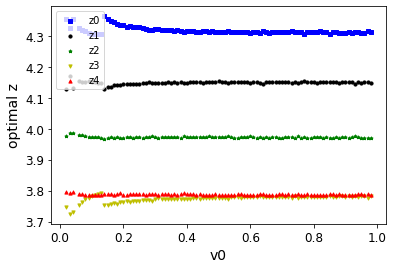}\label{dg_07_c}}
  \caption{Impact of varying vulnerability probability $v_0$ on the optimal $z^{*}$ with (a) $\alpha$'s = 0.1 (b) $\alpha$'s = 0.7}
\end{figure*}

The effect of network topology is studied for a dense and sparse graph with $N=5$.
    Fig.~\ref{dense_graph} and~\ref{sparse_graph} refers to the dense and sparse graph scenario considered for the Two-Stage model. Figs.~\ref{dg_01_c} and~\ref{dg_07_c} show the impact of varying  vulnerability probability $v_0$ on the optimal $z^{*}$ for $\alpha$'s = 0.1 and $\alpha$'s = 0.7, respectively. It can be observed that for lower $\alpha$ the optimal solutions have higher variance. For the $\alpha$ = 0.7 case, in comparison to the \textbf{simple model} (Fig.~\ref{dg_08}), there is no rise of $z_0$ with increasing $v_0$, indicating the reduction of sensitivity of optimal investment to updated $v_0$. 
    
\begin{figure*}[!htb]
  \centering
  \subfigure[]{\includegraphics[width=0.32\textwidth]{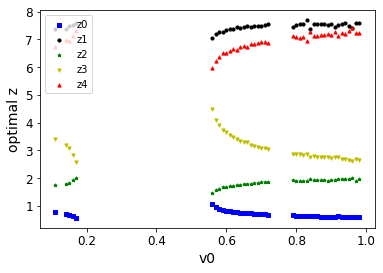}\label{sg_01_c}}
  \subfigure[]{\includegraphics[width=0.32\textwidth]{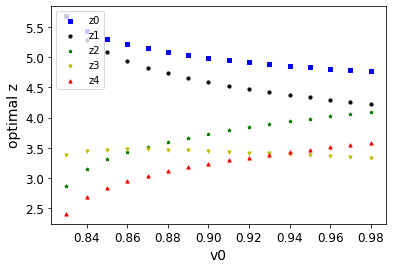}\label{sg_07_c}}
  \caption{Impact of varying vulnerability probability $v_0$ on the optimal $z^{*}$ with  a) $\alpha$'s = 0.1 (b) $\alpha$'s = 0.7}
\end{figure*}

    Figs.~\ref{sg_01_c} and ~\ref{sg_07_c} are the impact of varying  vulnerability probability $v_0$ on the optimal $z^{*}$ for $\alpha$'s = 0.1 and $\alpha$'s = 0.8 respectively. It can be observed that for lower $\alpha$ there exist multiple optimal solution in varying ranges of $v_0$, while with increase in $\alpha$ the range of $v_0$ for which an optimal solution exist reduced within 0.84 to 1.0 (Fig.~\ref{sg_07_c}). 
    
\section{Conclusion}\label{conclusion}
The vulnerability propagation and defense model is proposed and validated for both Two-node and N-node
systems. Sensitivity of the optimal investment with respect to the propagation factor and initial vulnerability probability is studied for both simple and two-stage models of the N-node scenarios with varying network topology. Finally, the proposed models are validated on a cyber-physical power system network comprising of utility control center and substation. The current work can be extended to dynamic vulnerability propagation and defense model with N-step integration which the current work demonstrates with a two-stage model. 
Moreover, the model can be validated in an emulated testbed such as Reslab~\cite{reslab}, where the $v$'s and $\alpha$'s can be estimated through various frameworks of cyber-physical situational awareness~\cite{framework}. 


\section{Acknowledgement}
The work was supported by funds from the US Department of Energy under award DE-OE0000895

{
\small
\bibliography{reference}
\bibliographystyle{IEEEtran}
}

\end{document}